# Architectural Blueprints—The "4+1" View Model of Software Architecture

*Philippe Kruchten*
Rational Software Corp.
638-650 West 41st Avenue
Vancouver, B.C., V5Z 2M9  Canada
pkruchten@rational.com

**Abstract**
This article presents a model for describing the architecture of software-intensive systems, based on the use of multiple, concurrent views. This use of multiple views allows to address separately the concerns of the various 'stakeholders' of the architecture: end-user, developers, systems engineers, project managers, etc., and to handle separately the functional and non functional requirements. Each of the five views is described, together with a notation to capture it. The views are designed using an architecture-centered, scenario-driven, iterative development process.

## Introduction

We all have seen many books and articles where one diagram attempts to capture the gist of the architecture of a system. But looking carefully at the set of boxes and arrows shown on these diagrams, it becomes clear that their authors have struggled hard to represent more on one blueprint than it can actually express. Are the boxes representing running programs? Or chunks of source code? Or physical computers? Or merely logical groupings of functionality? Are the arrows representing compilation dependencies? Or control flows? Or data flows? Usually it is a bit of everything. Does an architecture need a single architectural style? Sometimes the architecture of the software suffers scars from a system design that went too far into prematurely partitioning the software, or from an over-emphasis on one aspect of software development: data engineering, or run-time efficiency, or development strategy and team organization. Often also the architecture does not address the concerns of all its "customers" (or "stakeholders" as they are called at USC). This problem has been noted by several authors: Garlan & Shaw[1], Abowd & Allen at CMU, Clements at the SEI. As a remedy, we propose to organize the description of a software architecture using several concurrent *views*, each one addressing one specific set of concerns.

## An Architectural Model

Software architecture deals with the design and implementation of the high-level structure of the software. It is the result of assembling a certain number of architectural elements in some well-chosen forms to satisfy the major functionality and performance requirements of the system, as well as some other, non-functional requirements such as reliability, scalability, portability, and availability. Perry and Wolfe put it very nicely in this formula2, modified by Boehm:

> Software architecture = {Elements, Forms, Rationale/Constraints}

Software architecture deals with abstraction, with decomposition and composition, with style and esthetics. To describe a software architecture, we use a model composed of multiple *views* or perspectives. In order to eventually address large and challenging architectures, the model we propose is made up of five main views (cf. fig. 1):

- The logical view, which is the object model of the design (when an object-oriented design method is used),
- the *process* view, which captures the concurrency and synchronization aspects of the design,
- the *physical* view, which describes the mapping(s) of the software onto the hardware and reflects its distributed aspect,

- the *development* view, which describes the static organization of the software in its development environment.

The description of an architecture—the decisions made—can be organized around these four views, and then illustrated by a few selected *use cases*, or *scenarios* which become a fifth view. The architecture is in fact partially evolved from these scenarios as we will see later.

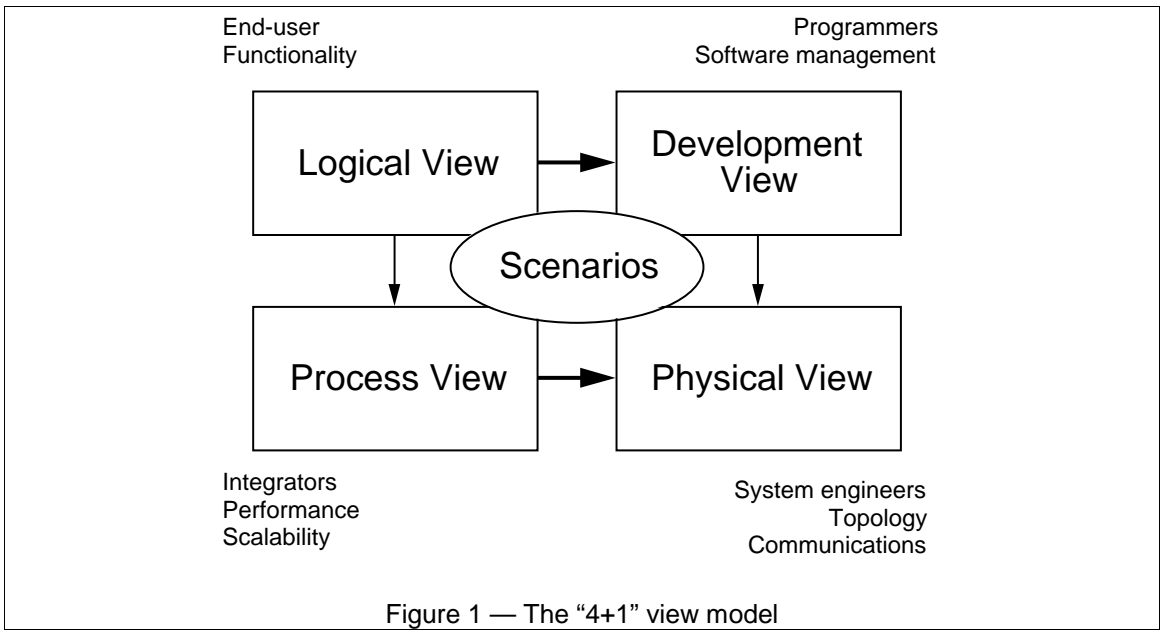

Figure 1 — The "4+1" view model

We apply Perry & Wolf's equation independently on each view, i.e., for each view we define the set of elements to use (components, containers, and connectors), we capture the forms and patterns that work, and we capture the rationale and constraints, connecting the architecture to some of the requirements.
Each view is described by a *blueprint* using its own particular notation. For each view also, the architects can pick a certain *architectural style*, hence allowing the coexistence of multiple styles in one system.

We will now look in turn at each of the five views, giving for each its purpose: which concerns is addresses, a notation for the corresponding architectural blueprint, the tools we have used to describe and manage it. Small examples are drawn from the design of a PABX, derived from our work at Alcatel Business System and an Air Traffic Control system[3], but in very simplified form—the intent here is just to give a flavor of the views and their notation and not to define the architecture of those systems.

The "4+1" view model is rather "generic": other notations and tools can be used, other design methods can be used, especially for the and the logical and process decompositions, but we have indicated the ones we have used with success.

# The Logical Architecture

*The Object-Oriented Decomposition*

The logical architecture primarily supports the functional requirements—what the system should provide in terms of services to its users. The system is decomposed into a set of key abstractions, taken (mostly) from the problem domain, in the form of *objects* or *object classes*. They exploit the principles of abstraction, encapsulation, and inheritance. This decomposition is not only for the sake of functional analysis, but also serves to identify common mechanisms and design elements across the various parts of the system. We use the Rational/Booch approach for representing the logical architecture, by means of *class diagrams* and *class templates*.[4] A class diagram shows a set of classes and their logical relationships: association, usage, composition, inheritance, and so forth. Sets of related classes can be grouped into class categories. Class templates focus on each individual class; they emphasize the main class operations, and identify key object characteristics. If it is important to define the internal behavior of an object, this is done with state transition diagrams, or state charts. Common mechanisms or services are defined in *class utilities*.

Alternatively to an OO approach, an application that is very data-driven may use some other form of logical view, such as E-R diagrams.

**Notation for the logical view**

The notation for the logical view is derived from the Booch notation[4]. It is considerably simplified to take into account only the items that are architecturally significant. In particular, the numerous adornments are not very useful at this level of design. We use Rational Rose® to support the logical architecture design.

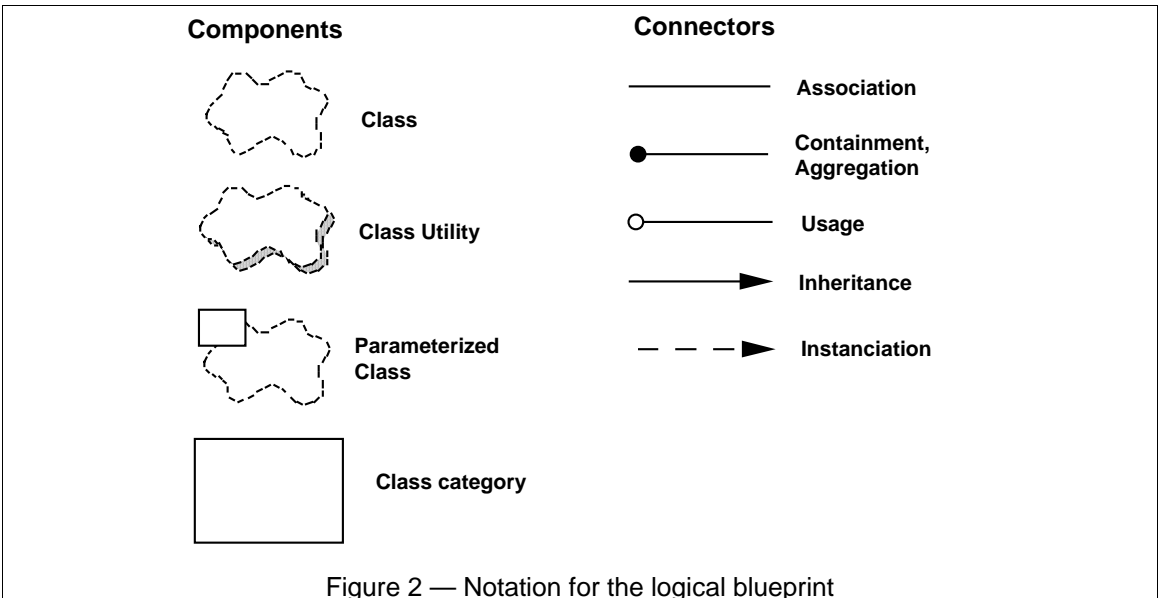

Figure 2 — Notation for the logical blueprint

**Style for the logical view**

The style we use for the logical view is an object-oriented style. The main guideline for the design of the logical view is to try to keep a single, coherent object model across the whole system, to avoid premature specialization of classes and mechanisms per site or per processor.

**Examples of Logical blueprints**
Figure 3a shows the main classes involved in the Télic PABX architecture.

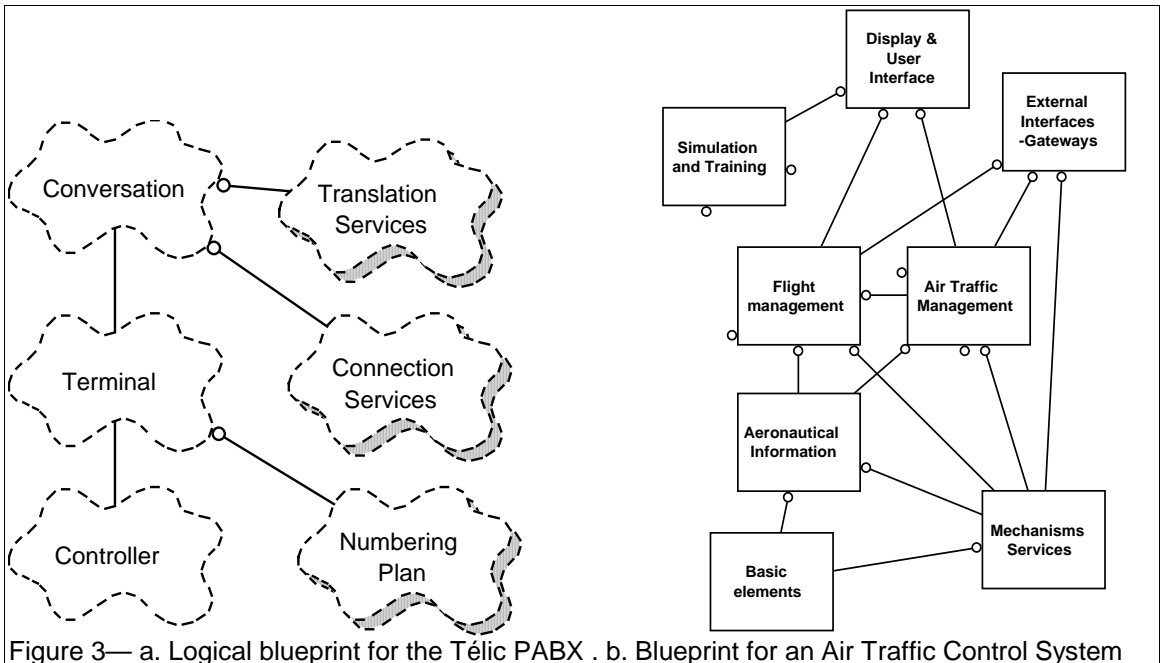

Figure 3— a. Logical blueprint for the Télic PABX . b. Blueprint for an Air Traffic Control System

A PABX establishes commmunications between terminals. A terminal may be a telephone set, a trunk line (i.e., line to central-office), a tie line (i.e., private PABX to PABX line), a feature phone line, a data line, an ISDN line, etc. Different lines are supported by different line interface cards. The responsibility of a line *controller* object is to decode and inject all the signals on the line interface card, translating card-specific signals to and from a small, uniform set of events: start, stop, digit, etc. The controller also bears all the hard real-time constraints. This class has many subclasses to cater for different kinds of interfaces. The responsibility of the *terminal* object is to maintain the state of a terminal, and negotiate services on behalf of that line. For example, it uses the services of the *numbering plan* to interpret the dialing in the selection phase. The *conversation* represents a set of terminals engaged in a conversation. The *conversation* uses *translation services* (directory, logical to physical address mapping, routes), and *connection* services to establish a voice path between the terminals.

For a much bigger system, which contains a few dozen classes of architectural significance, figure 3b show the top level class diagram of an air traffic control system, containing 8 class categories (i.e., groups of classes).

# The Process Architecture

*The Process Decomposition*
The process architecture takes into account some non-functional requirements, such as performance and availability. It addresses issues of concurrency and distribution, of system's integrity, of fault-tolerance, and how the main abstractions from the logical view fit within the process architecture—on which thread of control is an operation for an object actually executed.

The process architecture can be described at several levels of abstraction, each level addressing different concerns. At the highest level, the process architecture can be viewed as a set of independently executing logical *networks* of communicating programs (called "processes"), distributed across a set of hardware resources connected by a LAN or a WAN. Multiple logical networks may exist simultaneously, sharing the same physical resources. For example, independent logical networks may be used to support separation of the on-line operational system from the off-line system, as well as supporting the coexistence of simulation or test versions of the software.

A *process* is a grouping of tasks that form an executable unit. Processes represent the level at which the process architecture can be tactically controlled (i.e., started, recovered, reconfigured, and shut down). In addition, processes can be replicated for increased distribution of the processing load, or for improved availability.

The software is partitioned into a set of independent *tasks*. A task is a separate thread of control, that can be scheduled individually on one processing node.

We can distinguish then: major *tasks*, that are the architectural elements that can be uniquely addressed and minor *tasks*, that are additional tasks introduced locally for implementation reasons (cyclical activities, buffering, time-outs, etc.). They can be implemented as Ada tasks for example, or light-weight threads. Major tasks communicate via a set of well-defined inter-task communication mechanisms: synchronous and asynchronous message-based communication services, remote procedure calls, event broadcast, etc. Minor tasks may communicate by rendezvous or shared memory. Major tasks shall not make assumptions about their collocation in the same process or processing node.

Flow of messages, process loads can be estimated based on the process blueprint. It is also possible to implement a "hollow" process architecture with dummy loads for the processes, and measure its performance on the target system, as described by Filarey et al. in their Eurocontrol experiment.

**Notation for the Process view**

The notation we use for the process view is expanded from the notation originally proposed by Booch for Ada tasking. Again the notation used focuses on the elements that are architecturally significant. (Fig. 4)

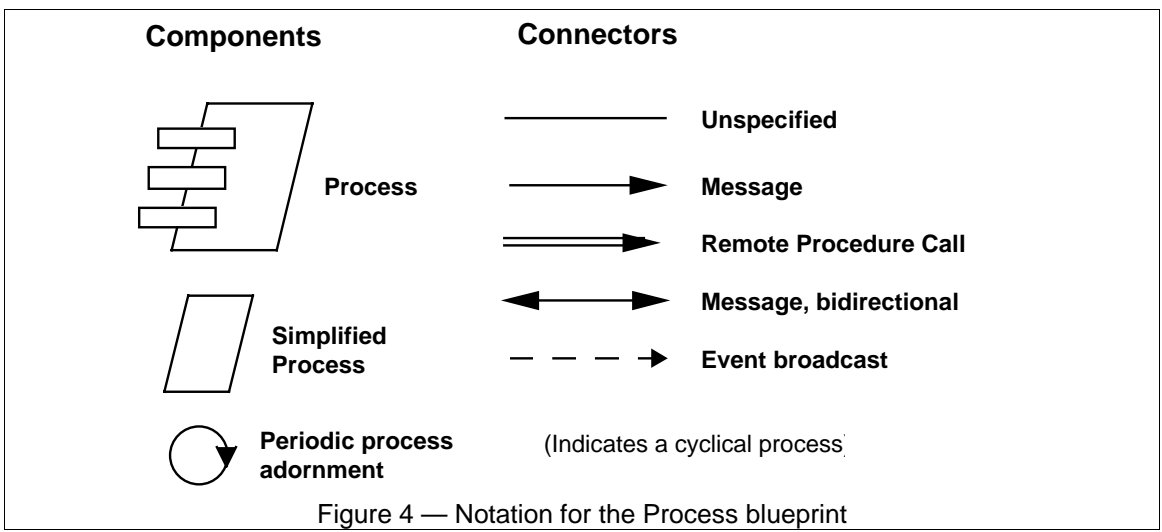

Figure 4 — Notation for the Process blueprint

We have used the Universal Network Architecture Services (UNAS) product from TRW to architect and implement the set of processes and tasks (and their redundancies) into networks of processes. UNAS contains a tool—the Software Architects Lifecycle Environment (SALE)—which supports such a notation. SALE allows for the graphical depiction of the process architecture, including specifications of the possible inter-task communication paths, from which the corresponding Ada or C++ source code is automatically generated. The benefit of this approach to specifying and implementing the process architecture is that changes can be incorporated easily without much impact on the application software.

**Style for the process view**

Several styles would fit the process view. For example, picking from Garlan and Shaw's taxonomy[1] we can have: pipes and filters, or client/server, with variants of multiple client/single server and multiple clients/multiple servers. For more complex systems, one could use a style similar to the process groups approach of the ISIS system as described by K. Birman with another notation and toolset.

**Example of a Process blueprint**

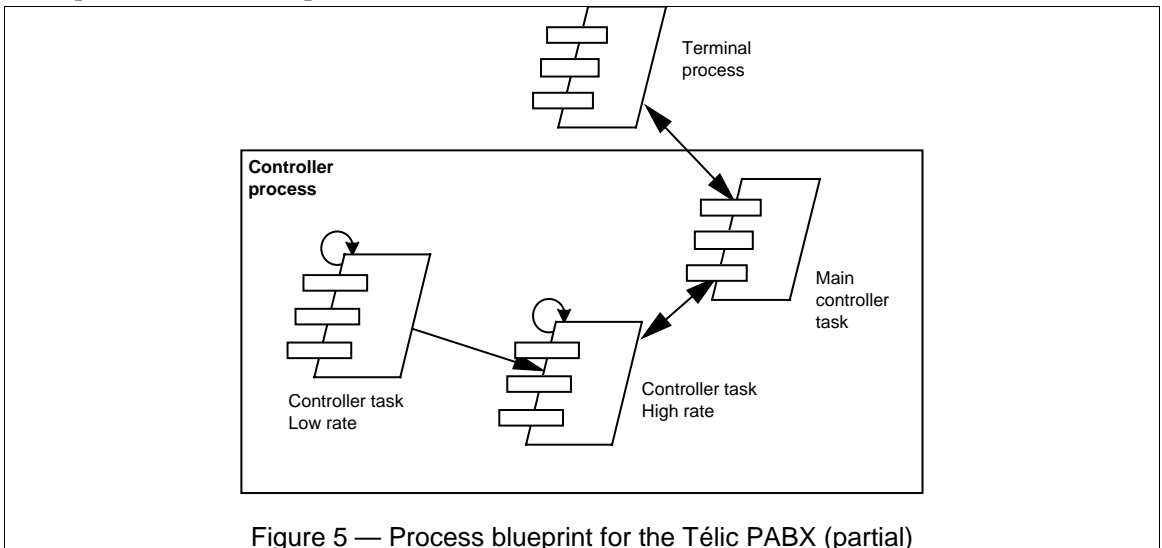

Figure 5 — Process blueprint for the Télic PABX (partial)

All terminals are handled by a single *terminal process*, which is driven by messages in its input queues. The controller objects are executed on one of three tasks that composes the controller process: a *low cycle rate task* scans all inactive terminals (200 ms), puts any terminal becoming active in the scan list of the *high cycle rate task* (10ms), which detects any significant change of state, and passes them to the *main controller task* which interprets the changes and communicates them by message to the corresponding terminal. Here message passing within the controller process is done via shared memory.

# The Development Architecture

*Subsystem decomposition*

The development architecture focuses on the actual software module organization on the software development environment. The software is packaged in small chunks—program libraries, or *subsystems*—that can be developed by one or a small number of developers. The subsystems are organized in a hierarchy of *layers*, each layer providing a narrow and well-defined interface to the layers above it.

The development architecture of the system is represented by module and subsystem diagrams, showing the 'export' and 'import' relationships. The complete development architecture can only be described when all the elements of the software have been identified. It is, however, possible to list the rules that govern the development architecture: partitioning, grouping, visibility.

For the most part, the development architecture takes into account internal requirements related to the ease of development, software management, reuse or commonality, and to the constraints imposed by the toolset, or the programming language. The development view serves as the basis for requirement allocation, for allocation of work to teams (or even for team organization), for cost evaluation and planning, for monitoring the progress of the project, for reasoning about software reuse, portability and security. It is the basis for establishing a line-of-product.

**Notation for the Development Blueprint**

Again, a variation of the Booch notation, limiting it to the items that are architecturally significant.

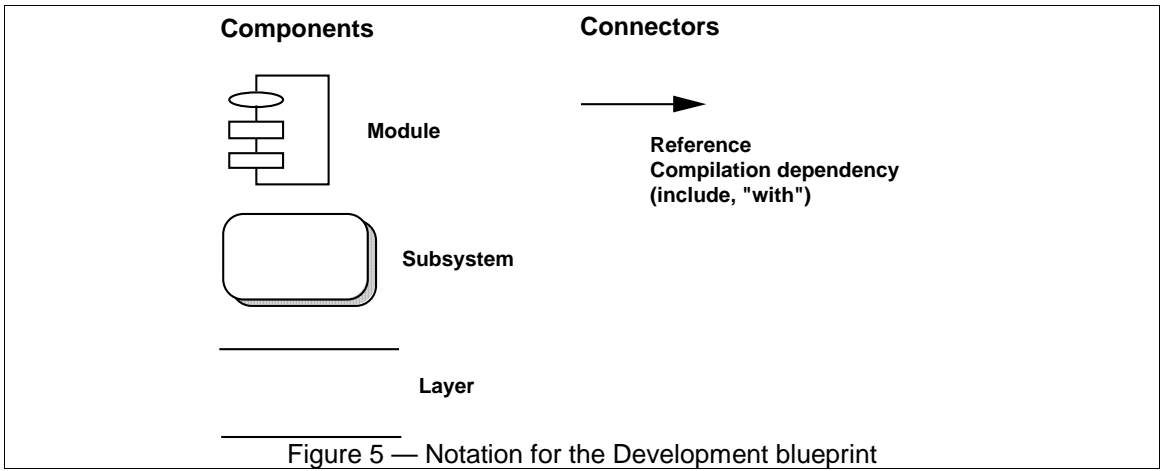
Figure 5 — Notation for the Development blueprint

The Apex Development Environment from Rational supports the definition and the implementation of the development architecture, the layering strategy described above, and the enforcement of the design rules. Rational Rose can draw the development blueprints at the module and subsystem level, in forward engineering and by reverse engineering from the development source code, for Ada and C++.

**Style for the Development View**

We recommend adopting a *layered style* for the development view, defining some 4 to 6 layers of subsystems. Each layer has a well-defined responsibility. The design rule is that a subsystem in a certain can only depend on subsystem that are in the same layer or in layers below, in order to minimize the development of very complex networks of dependencies between modules and allow simple release strategies layer by layer.

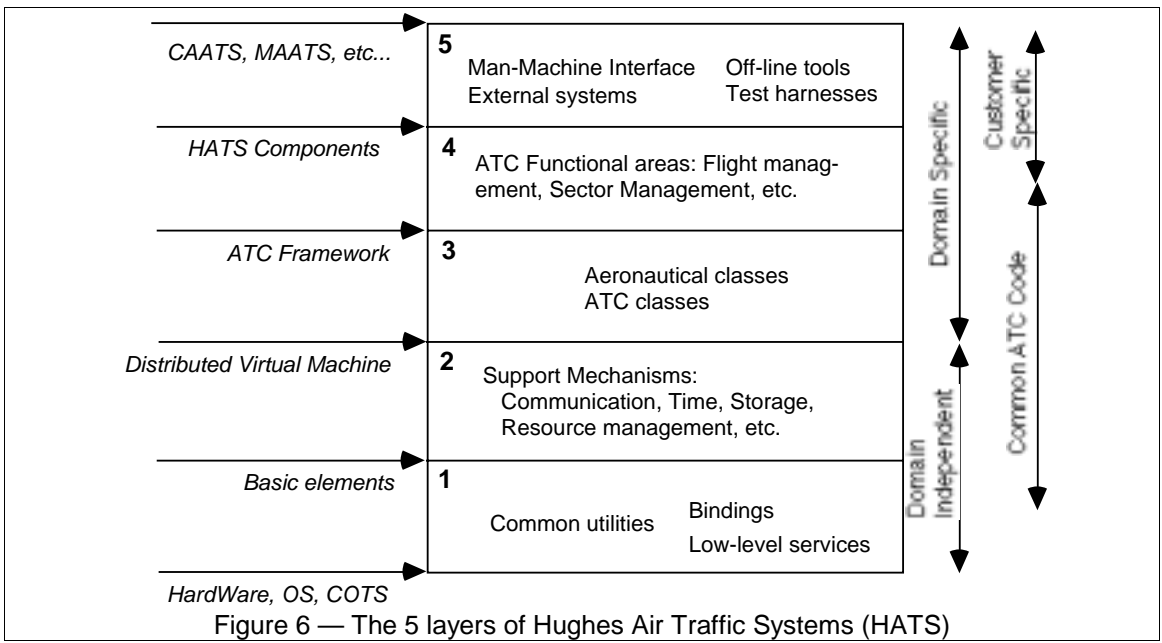
Figure 6 — The 5 layers of Hughes Air Traffic Systems (HATS)

**Example of Development architecture**

Figure 6 represents the development organization in five layers of a line-of-product of Air Traffic Control systems developed by Hughes Aircraft of Canada[3]. This is the development architecture corresponding to the logical architecture shown in fig. 3b.

Layers 1 and 2 constitute a domain-independent distributed infrastructure that is common across the line of products and shields it from variations in hardware platform, operating system, or off-the-shelf products such as database management system. To this infrastructure, layer 3 adds an ATC framework to form a *domain-specific software architecture*. Using this framework a palette of functionality is build in layer 4. Layer 5 is very customer- and product-dependent, and contains most of the user-interface and interfaces with the external systems. Some 72 subsystems are spread across of the 5 layers, containing each from 10 to 50 modules, and can be represented on additional blueprints.

# The Physical Architecture

*Mapping the software to the hardware*

The physical architecture takes into account primarily the non-functional requirements of the system such as availability, reliability (fault-tolerance), performance (throughput), and scalability. The software executes on a network of computers, or processing nodes (or just *nodes* for short). The various elements identified—networks, processes, tasks, and objects—need to be mapped onto the various nodes. We expect that several different physical configurations will be used: some for development and testing, others for the deployment of the system for various sites or for different customers. The mapping of the software to the nodes therefore needs to be highly flexible and have a minimal impact on the source code itself.

**Notation for the Physical Blueprint**

Physical blueprints can become very messy in large systems, so they take several forms, with or without the mapping from the process view.

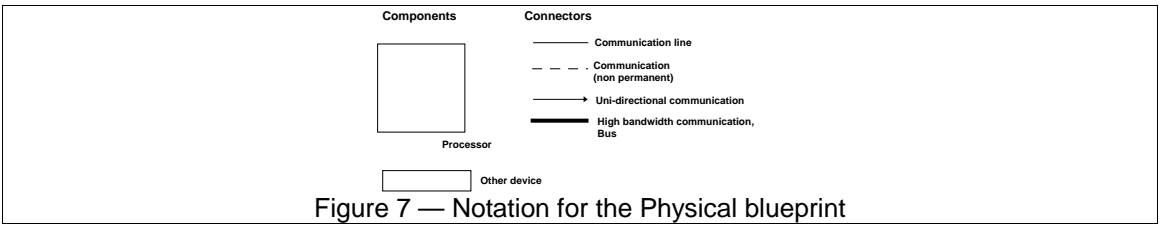

Figure 7 — Notation for the Physical blueprint

UNAS from TRW provide us here with data-driven means of mapping the process architecture onto the physical architecture allowing a large class of changes in the mapping without source code modifications.

**Example of Physical blueprint**

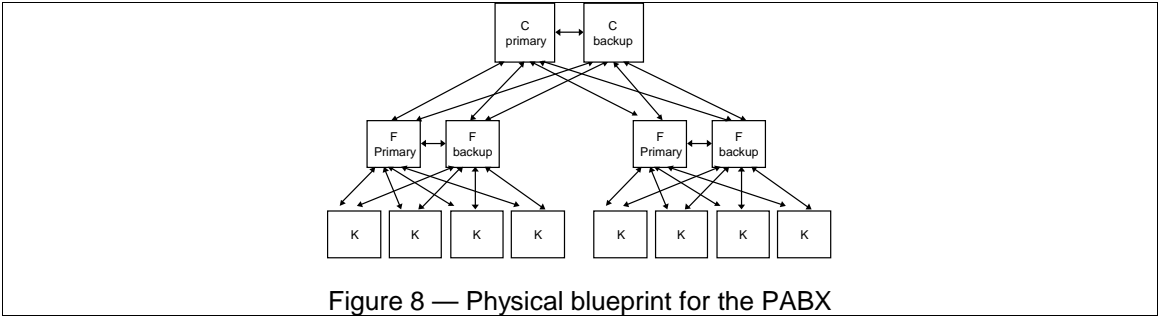

Figure 8 — Physical blueprint for the PABX

Figure 8 shows one possible hardware configuration for a large PABX, whereas figures 9 and 10 show mappings of the process architecture on two different physical architectures, corresponding to a small and a large PABX. C, F and K are three types of computers of different capacity, supporting three different executables.

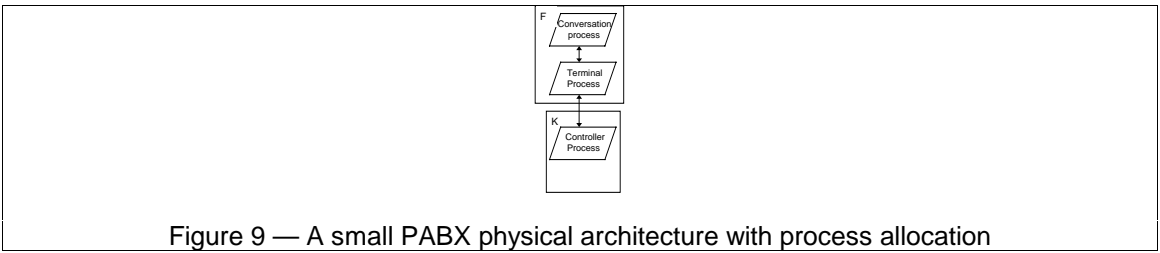

Figure 9 — A small PABX physical architecture with process allocation

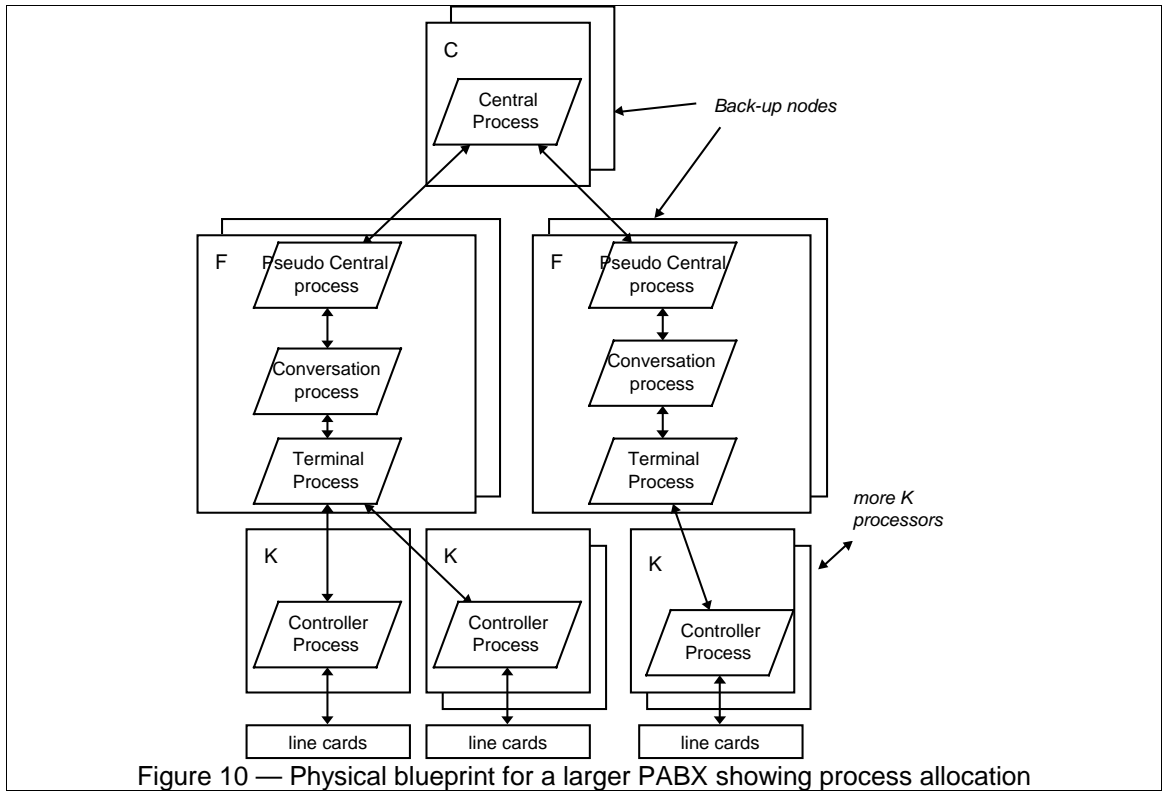
Figure 10 — Physical blueprint for a larger PABX showing process allocation

# Scenarios

*Putting it all together*

The elements in the four views are shown to work together seamlessly by the use of a small set of important *scenarios* —instances of more general *use cases*—for which we describe the corresponding scripts (sequences of interactions between objects, and between processes) as described by Rubin and Goldberg[6]. The scenarios are in some sense an abstraction of the most important requirements. Their design is expressed using object scenario diagrams and object interaction diagrams[4].

This view is redundant with the other ones (hence the "+1"), but it serves two main purposes:
- as a driver to discover the architectural elements during the architecture design as we will describe later
- as a validation and illustration role after this architecture design is complete, both on paper and as the starting point for the tests of an architectural prototype.

**Notation for the Scenarios**

The notation is very similar to the Logical view for the components (cf. fig. 2), but uses the connectors of the Process view for interactions between objects (cf. fig. 4). Note that object instances are denoted with solid lines. As for the logical blueprint, we capture and manage object scenario diagrams using Rational Rose.

**Example of a Scenario**

Fig. 11 shows a fragment of a scenario for the small PABX. The corresponding *script* reads:
1. The controller of Joe's phone detects and validate the transition from on-hook to off-hook and sends a message to wake up the corresponding terminal object.
2. The terminal allocates some resources, and tells the controller to emit some dial-tone.
3. The controller receives digits and transmits them to the terminal.
4. The terminal uses the numbering plan to analyze the digit flow.
5. When a valid sequence of digits has been entered, the terminal opens a conversation.

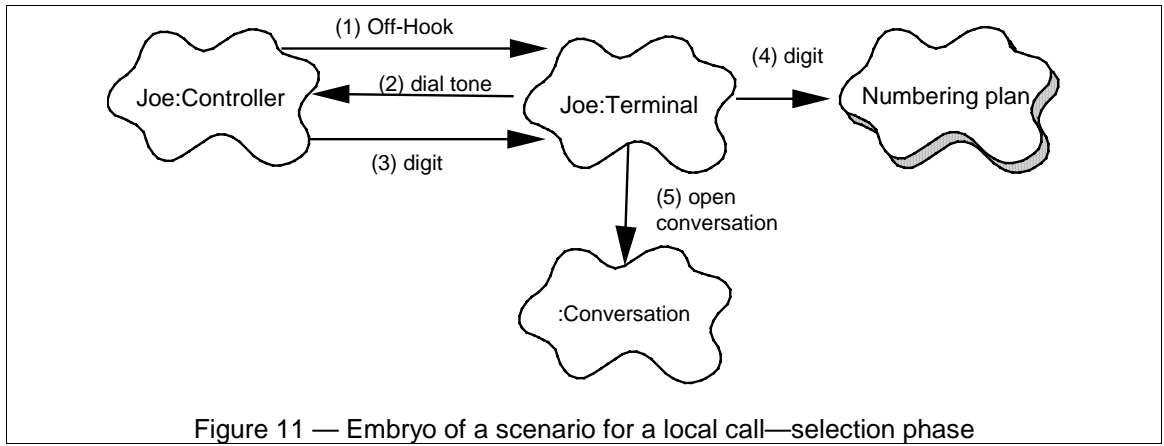

Figure 11 — Embryo of a scenario for a local call—selection phase

## Correspondence Between the Views

The various views are not fully orthogonal or independent. Elements of one view are connected to elements in other views, following certain design rules and heuristics.

**From the logical to the process view**

We identify several important characteristics of the classes of the logical architecture:
- Autonomy: are the objects active, passive, protected?
  -an *active* object takes the initiative of invoking other objects' operations or its own operations, and has full control over the invocation of its own operations by other objects
  -a *passive* object never invokes spontaneously any operations and has no control over the invocation of its own operations by other objects
  - a *protected* object never invokes spontaneously any operations but performs some arbitration on the invocation of its operations.
- Persistence: are the objects transient , permanent? Do they the failure of a process or processor?
- Subordination: are the existence or persistence of an object depending on another object?
- Distribution: are the state or the operations of an object accessible from many nodes in the physical architecture, from several processes in the process architecture?

In the logical view of the architecture we consider each object as active, and potentially "concurrent," i.e., behaving "in parallel" with other objects, and we pay no more attention to the exact degree of concurrency we need to achieve this effect. Hence the logical architecture takes into account only the functional aspect of the requirements.

However when we come to defining the process architecture, implementing each object with its own thread of control (e.g., its own Unix process or Ada task) is not quite practical in the current state of technology, because of the huge overhead this imposes. Moreover, if objects are concurrent, there must be some form of arbitration for invoking their operations.

On another hand, multiple threads of control are needed for several reasons:
- To react rapidly to certain classes of external stimuli, including time-related events
- To take advantage of multiple CPUs in a node, or multiple nodes in a distributed system
- To increase the CPU utilization, by allocating the CPU to other activities while some thread of control is suspended waiting for some other activity to complete (e.g., access to some external device, or access to some other active object)
- To prioritize activities (and potentially improve responsiveness)
- To support system scalability (with additional processes sharing the load)
- To separate concerns between different areas of the software
- To achieve a higher system availability (with backup processes)

We use concurrently two strategies to determine the 'right' amount of concurrency and define the set of processes that are needed. Keeping in mind the set of potential physical target architectures, we can proceed either:
- **Inside-out:**
  Starting from the logical architecture: define agent tasks which multiplex a single thread of control across multiple active objects of a class; objects whose persistency or life is subordinate to an active object are also executed on that same agent; several classes that need to be executed in mutual exclusion, or that require only small amount of processing share a single agent. This clustering proceeds until we have reduced the processes to a reasonably small number that still allows distribution and use of the physical resources.

- **Outside-in:**
  Starting with the physical architecture: identify external stimuli (requests) to the system, define client processes to handle the stimuli and servers processes that only provide services and do not initiate them; use the data integrity and serialization constraints of the problem to define the right set of servers, and allocate objects to the client and servers agents; identify which objects must be distributed.

The result is a mapping of classes (and their objects) onto a set of tasks and processes of the process architecture. Typically, there is an *agent* task for an active class, with some variations: several agents for a given class to increase throughput, or several classes mapped onto a single agent because their operations are infrequently invoked or to guarantee sequential execution.

Note that this is not a linear, deterministic process leading to an optimal process architecture; its requires a few iterations to get an acceptable *compromise*. There are numerous other ways to proceed, as shown by Birman et al.[5] or Witt et al.[7] for example. The precise method used to construct the mapping is outside of the scope of this article, but we can illustrate it on a small example.

Fig. 12 shows how a small set of classes from some hypothetical air-traffic control system maybe mapped onto processes.

The *flight class* is mapped onto a set of *flight agents*: there are many flights to process, a high rate of external stimuli, response time is critical, the load must be spread across multiple CPUs. Moreover the persistency and distribution aspects of the flight processing are deferred to a *flight server*, which is duplicated for availability reasons.

A flight *profile* or a *clearance* are always subordinate to a flight, and although there are complex classes, they share the processes of the flight class. Flights are distributed to several other processes, notably for to display and external interfaces.

A *sectorization class*, which established a partitioning of airspace for the assignment of jurisdiction of controllers over flights, because of its integrity constraints, can be handled only by a single agent, but can share the server process with the flight: updates are infrequent.

*Locations* and *airspace* and other static aeronautical information are protected objects, shared among several classes, rarely updated; they are mapped on their own server, and distributed to other processes.

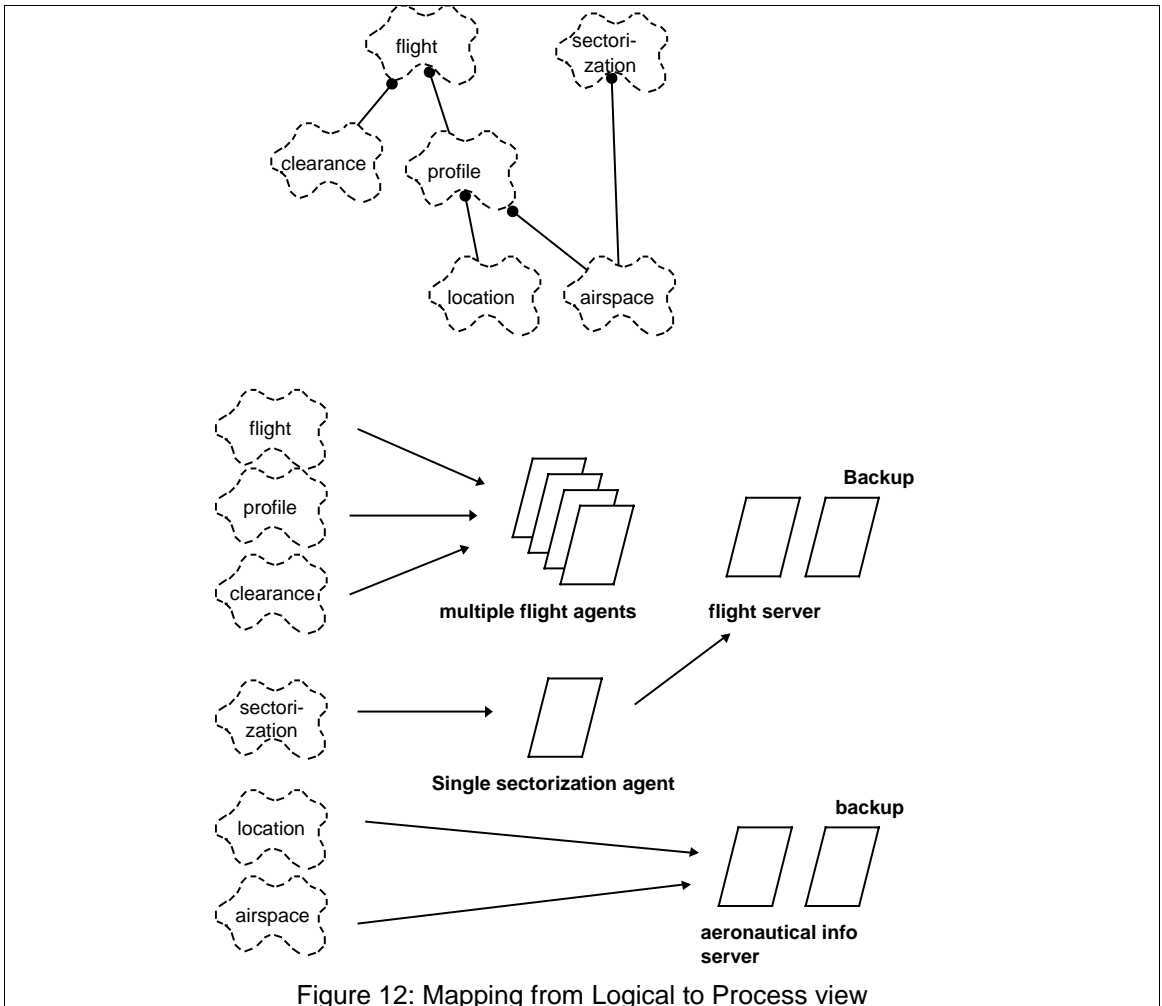

Figure 12: Mapping from Logical to Process view

**From logical to development**

A class is usually implemented as a module, for example a type in the visible part of an Ada *package*. Large classes are decomposed into multiple packages. Collections of closely related classes—class categories—are grouped into subsystems. Additional constraints must be considered for the definition of subsystems, such as team organization, expected magnitude of code (typically 5K to 20K SLOC per subsystem), degree of expected reuse and commonality, and strict layering principles (visibility issues), release policy and configuration management. Therefore we usually end up with a view that does not have a one to one correspondence with the logical view.

The logical and development views are very close, but address very different concerns. We have found that the larger the project, the greater the distance between these views. Similarly for the process and physical views: the larger the project, the greater the distance between the views. For example, if we compare fig. 3b and fig. 6, there is no one to one mapping of the class categories to the layers. If we take the 'External interfaces—Gateway' category, its implementation is spread across several layers: communications protocols are in subsystems in or below layer 1, general gateway mechanisms are in subsystems in layer 2, and the actual specific gateways in layer 5 subsystems.

**From process to physical**

Processes and process groups are mapped onto the available physical hardware, in various configurations for testing or deployment. Birman describes some very elaborate schemes for this mapping in the Isis project[5].

The scenarios relate mostly to the logical view, in terms of which classes are used, and to the process view when the interactions between objects involve more than one thread of control.

## Tailoring the Model

Not all software architecture need the full "4+1" views. Views that are useless can be omitted from the architecture description, such as the physical view, if there is only one processor, and the process view if there is only process or program. For very small system, it is even possible that the logical view and the development view are so similar that they do not require separate descriptions. The scenarios are useful in all circumstances.

## Iterative process

Witt et al. indicate 4 phases for the design or an architecture: sketching, organizing, specifying and optimizing, subdivided into some 12 steps[7]. They indicate that some backtracking may be needed. We think that this approach is too "linear" for an ambitious and rather unprecedented project. Too little is known at the end of the 4 phases to validate the architecture. We advocate a more iterative development, were the architecture is actually prototyped, tested, measured, analyzed, and then refined in subsequent iterations. Besides allowing to mitigate the risks associated with the architecture, such an approach has other side benefits for the project: team building, training, acquaintance with the architecture, acquisition of tools, run-in of procedures and tools, etc. (We are speaking here of an evolutionary prototype, that slowly grows into becoming the system, and not of throw-away, exploratory prototypes.) This iterative approach also allows the requirements to be refined, matured, better understood.

**A scenario-driven approach**

The most critical functionality of the system is captured in the form of scenarios (or use cases). By critical we mean: functions that are the most important, the *raison d'être* of the system, or that have the highest frequency of use, or that present some significant technical risk that must be mitigated.
*Start:*
- A small number of the scenarios are chosen for an iteration based on risk and criticality. Scenarios may be synthesized to abstract a number of user requirements.
- A strawman architecture is put in place. The scenarios are then "scripted" in order to identify major abstractions (classes, mechanisms, processes, subsystems) as indicated by Rubin and Goldberg[6] — decomposed in sequences of pairs (object, operation).
- The architectural elements discovered are laid out on the 4 blueprints: logical, process, development, and physical.
- This architecture is then implemented, tested, measured, and this analysis may detect some flaws or potential enhancement.
- Lessons learned are captured.

*Loop:*

 The next iteration can then start by:
- reassessing the risks,
- extending the palette of scenarios to consider
- selecting a few additional scenarios that will allow risk mitigation or greater architecture coverage

 Then:
- Try to script those scenarios in the preliminary architecture
- discover additional architectural elements, or sometimes significant architectural changes that need to occur to accommodate these scenarios
- update the 4 main blueprints: logical, process, development, physical
- revise the existing scenarios based on the changes
- upgrade the implementation (the architectural prototype) to support the new extended set of scenario.
- Test. Measure under load, in real target environment if possible.
- All five blueprints are then reviewed to detect potential for simplification, reuse, commonality.
- Design guidelines and rationale are updated.
- Capture the lessons learned.

*End loop*

The initial architectural prototype evolves to become the real system. Hopefully after 2 or 3 iterations, the architecture itself become stable: no new major abstractions are found, no new subsystems or processes, no new interfaces. The rest of the story is in the realm of software design, where, by the way, development may continue using very similar methods and process.

The duration of these iterations varies considerably: with the *size* of the project to put in place, with the *number of people* involved and their familiarity with the domain and with the method, and with the *degree of "unprecedentedness"* of the system w.r.t. this development organization. Hence the duration of an iteration may be 2-3 weeks for a small project (e.g., 10 KSLOC), or up to 6-9 months for a large command and control system (e.g., 700 KSLOC).

## Documenting the architecture

The documentation produced during the architectural design is captured in two documents:
- A *Software Architecture Document*, whose organization follows closely the "4+1" views (cf. fig. 13 for a typical outline)
- A *Software Design Guidelines*, which captures (among other things) the most important design decisions that must be respected to maintain the architectural integrity of the system.

```
Title Page
Change History
Table of Contents
List of Figures
1. Scope
2. References
3. Software Architecture
4. Architectural Goals & Constraints
5. Logical Architecture
6. Process Architecture
7. Development Architecture
8. Physical Architecture
9. Scenarios
10. Size and Performance
11. Quality
Appendices
   A. Acronyms and Abbreviations
   B. Definitions
   C. Design Principles
```

Figure 13 — Outline of a Software Architecture Document

## Conclusion

This "4+1" view model has been used with success on several large projects with or without some local customization and adjustment in terminology[4]. It actually allowed the various stakeholders to find what they want to know about the software architecture. Systems engineers approach it from the Physical view, then the Process view. End-users, customers, data specialists from the Logical view. Project managers, software configuration staff see it from the Development view.

Other sets of views have been proposed and discussed, within Rational and elsewhere, for instance by Meszaros (BNR), Hofmeister, Nord and Soni (Siemens), Emery and Hilliard (Mitre)[8], but we have found that often these other views proposed could usually be folded into one of the 4 we described. For example a Cost & Schedule view folds into the Development view, a Data view into the Logical view, an Execution view into a combination of the Process and Physical view.

| *View* | *Logical* | *Process* | *Development* | *Physical* | *Scenarios* |
|---|---|---|---|---|---|
| *Components* | Class | Task | Module, Subsystem | Node | Step, Scripts |
| *Connectors* | association, inheritance, containment | Rendez-vous, Message, broadcast, RPC, etc. | compilation dependency, "with" clause, "include" | Communication medium, LAN, WAN, bus, etc. | |
| *Containers* | Class category | Process | Subsystem (library) | Physical subsystem | Web |

| *Stakeholders* | End-user | System designer, integrator | Developer, manager | System designer | End-user, developer |
| --- | --- | --- | --- | --- | --- |
| *Concerns* | Functionality | Performance, availability, S/W fault-tolerance, integrity | Organization, reuse, portability, line-of-product | Scalability, performance, availability | Understand-ability |
| *Tool support* | Rose | UNAS/SALE DADS | Apex, SoDA | UNAS, Openview DADS | Rose |

Table 1 — Summary of the "4+1" view model

**Acknowledgments**

The "4+1" view model owes its existence to many colleagues at Rational, at Hughes Aircraft of Canada, at Alcatel, and elsewhere. In particular I would like to thank for their contributions Ch. Thompson, A. Bell, M. Devlin, G. Booch, W. Royce, J. Marasco, R. Reitman, V. Ohnjec, and E. Schonberg.## References